# A fifth order perturbative solution to the Gribov–Lipatov equation[1]

**Mariusz Przybycień**

Deutsches Elektronen–Synchrotron DESY
Notkestrasse 85, 2000 Hamburg 52, FRG
and
Institute of Physics, Jagellonian University,
Reymonta 4, 30–059 Cracow, Poland

**Abstract**

Fifth order, $\mathcal{O}(\beta^5)$, exact corrections to the non–singlet electron structure function in QED are presented. Calculations were performed in the leading logarithmic approximation using the *ad hoc* exponentiation prescription proposed by Jadach and Ward and a recurence formula for the elements of the Jadach–Ward series. A comparison with existing third order, $\mathcal{O}(\beta^3)$, solutions is also presented. The three next elements of the Jadach–Ward series were calculated numerically and parametrized with an accuracy better than $5 \cdot 10^{-6}$ in the range of x between 0.01 and 1.

---

[1] Work partly supported by the KBN Grant No. 20-38-09101

# 1 Introduction

The leading logarithmic non–singlet electron structure function $D^{NS}(x,\beta)$ is defined as the difference between the densities of virtual electrons and positrons in the initial electron. It can be found by solving the Gribov–Lipatov evolution equation [1]:

$$D^{NS}(x,\beta) = \delta(1-x) + \frac{1}{4}\int_0^\beta d\eta D^{NS}(\cdot,\eta) \otimes P(\cdot)(x) \qquad (1)$$

where:

$$\beta(s) = \frac{2}{\pi}\int_{m_e^2}^s \frac{ds'}{s'}\alpha(s') \qquad (2)$$

and

$$P(z) = \delta(1-z)\left(\frac{3}{2} + 2\ln\epsilon\right) + \Theta(1-\epsilon-z)\frac{1+z^2}{1-z}. \qquad (3)$$

In equation (2) $\alpha(s)$ is the running coupling constant and $s = Q^2$ is the energy scale characteristic for the process. In the case of $\alpha(s) = \alpha$ we find that:

$$\beta(s) = \frac{2\alpha}{\pi}\ln\frac{s}{m_e^2}. \qquad (4)$$

The convolution symbol $\otimes$ stands for:

$$P_1(\cdot) \otimes P_2(\cdot)(x) = \int_0^1\int_0^1 dx_1 dx_2 \delta(x - x_1 x_2) P_1(x_1) P_2(x_2). \qquad (5)$$

We can rewrite equation (1) in the differential form:

$$\frac{\partial D^{NS}(x,\beta)}{\partial \beta} = \frac{1}{4}\int_0^1\int_0^1 dx_1 dx_2 \delta(x - x_1 x_2) P(x_1) D^{NS}(x_2,\beta) \qquad (6)$$

with the boundary condition

$$D^{NS}(x,0) = \delta(1-x) \qquad (7)$$

which means that for $s = m_e^2$ the electron does not have any internal structure. More information about leading logarithmic calculations of the QED corrections can be found in [2]. In reference [2] the perturbative solution to the equations (6)-(7) up to third order in $\beta$ has been calculated for various prescriptions of *ad hoc* exponentiations.

The main result obtained in this paper is the exact solution to the above equation up to the fifth order in $\beta$. The accuracy of the solution is estimated and the comparison with the existing third order solutions is presented. I propose also an approximate solution up to the eighth order in $\beta$ valid in the range of 0.01< $x$ <1 with accuracy of the order of $10^{-6}$. An interesting symmetry property of the solution is mentioned.

## 2 Solution to the GL equation

According to the prescription of the *ad hoc* exponentiation procedure proposed by Jadach and Ward [3] we look for the solution to the Gribov–Lipatov equation in the following form:

$$D_{JW}^{NS}(x,\beta) = D_G(x,\beta) \sum_{n=0}^{N-1} \beta^n \phi_n(x) + \mathcal{O}(\beta^{N+1}) \tag{8}$$

where the Gribov function reads:

$$D_G(x,\beta) = \frac{exp[\beta/2(3/4-\gamma)]}{\Gamma(1+\beta/2)} \frac{\beta}{2} (1-x)^{\beta/2-1} \tag{9}$$

and is a solution of equation (6) for $x \to 1^-$. At the beginning we eliminate the IR cut-off $\epsilon$ and arrive at the following evolution equation:

$$\begin{aligned}
\frac{\partial D^{NS}(x,\beta)}{\partial \beta} &= \left[\frac{3}{8} + \frac{1}{2}\ln(1-x) - \frac{1}{2}\ln x\right] D^{NS}(x,\beta) \\
&+ \frac{1}{4}\int_x^1 \frac{dy}{y-x}\left[\left(1+\frac{x^2}{y^2}\right)D^{NS}(y,\beta) - 2D^{NS}(x,\beta)\right].
\end{aligned} \tag{10}$$

Substituting:

$$D_{JW}^{NS}(x,\beta) = D_G(x,\beta)\Phi(x,\beta), \tag{11}$$

we get for $\Phi(x,\beta)$:

$$\begin{aligned}
\frac{\partial \Phi(x,\beta)}{\partial \beta} &= \frac{1}{\beta}\left[\frac{1}{2}(1+x^2) - \Phi(x,\beta)\right] - \frac{1}{2}\ln x \Phi(x,\beta) \\
&+ \frac{1}{4}\int_x^1 dy \left(\frac{1-y}{1-x}\right)^{\frac{\beta}{2}} \left\{\frac{1}{y-x}\left[\frac{1-x}{1-y}\left(1+\frac{x^2}{y^2}\right)\Phi(y,\beta) - 2\Phi(x,\beta)\right] - \frac{1+x^2}{1-y}\right\}
\end{aligned} \tag{12}$$

with the condition

$$\Phi(1,\beta) = 1. \tag{13}$$

We can write the solution of the above equation in terms of a power series in $\beta$. Substituting:

$$\Phi(x,\beta) = \sum_{n=0}^{\infty} (\beta/2)^n \phi_n(x) \tag{14}$$

one can derive a recurence formula for the coefficient functions of the Jadach–Ward series:

$$\phi_0(x) = \frac{1}{2}(1+x^2) \tag{15}$$

$$\phi_1(x) = -\frac{1}{8}[2(1-x)^2 + (1+3x^2)\ln x] \tag{16}$$

$$\begin{aligned}
\phi_{n+1}(x) &= \frac{1}{n+2}\left\{\frac{1}{4}(1-x)\lambda_n(x) - \phi_n(x)\ln x + \frac{1}{2}(1-x)\sum_{k=1}^{n}\frac{1}{(n-k)!}\right. \\
&\left. \int_x^1 \frac{dy}{y-x}\ln^{n-k}\left(\frac{1-y}{1-x}\right)\left[\left(1+\frac{x^2}{y^2}\right)\frac{\phi_k(y)}{1-y} - 2\frac{\phi_k(x)}{1-x}\right]\right\}
\end{aligned} \tag{17}$$

where
$$\lambda_n(x) = -\frac{1}{n!} \int_x^1 \frac{dy}{y^2} (1+y)(x+y) \ln^n \left( \frac{1-y}{1-x} \right). \tag{18}$$

The above recurence formula was first obtained in [4]. $\lambda_n(x)$ can be expressed in terms of Nielsen's polilogarithms:

$$\lambda_1(x) = 1 - x + (1+x)Li_2(1-x) - x \ln x \tag{19}$$

and for $n \geq 2$

$$\lambda_n(x) = (-1)^{n+1}[1 - x + (1+x)S_{n,1}(1-x) + xS_{n-1,1}(1-x)] \tag{20}$$

Some of the definitions and formulas concerning the Nielsen's polilogarithms useful in following calculations are collected in Appendix A. Using the recurence formula (17) one can calculate a few next coefficients of the Jadach–Ward series:

$$\phi_2(x) = \frac{1}{8} \left[ (1-x)^2 + \frac{1}{2}(1 - 4x + 3x^2) \ln x + \frac{1}{12}(1 + 7x^2) \ln^2 x + (1 - x^2) Li_2(1-x) \right] \tag{21}$$

$$\phi_3(x) = -\frac{1}{96} \left[ 5(1-x)^2 + \frac{1}{2}(5 - 24x + 19x^2) \ln x + \frac{1}{2}(1 - 8x + 7x^2) \ln^2 x \right.$$
$$+ \frac{1}{24}(1 + 15x^2) \ln^3 x + \left( 6(1-x^2) + 4(1+x^2) \ln x \right) Li_2(1-x)$$
$$\left. + 12(1-x^2) Li_3(1-x) + 2(1 + 7x^2) S_{1,2}(1-x) \right] \tag{22}$$

$$\phi_4(x) = \frac{1}{384} \left[ 7(1-x)^2 + \frac{1}{2}(7 - 40x + 33x^2) \ln x + \frac{1}{4}(3 - 34x + 31x^2) \ln^2 x \right.$$
$$+ \frac{1}{12}(1 - 16x + 15x^2) \ln^3 x + \frac{1}{240}(1 + 31x^2) \ln^4 x + 8(1+x^2) Li_2^2(1-x)$$
$$+ \left( 14(1-x^2) + 8(1-x)^2 \ln x + \frac{1}{2}(3 + 13x^2) \ln^2 x \right) Li_2(1-x)$$
$$+ \left( 24(1-x^2) + 16(1+x^2) \ln x \right) Li_3(1-x) + 48(1-x^2) Li_4(1-x)$$
$$+ \left( 4(1 - 8x + 7x^2) - (1 - 17x^2) \ln x \right) S_{1,2}(1-x)$$
$$\left. - 24(1-x^2) S_{2,2}(1-x) - 5(1-x^2) S_{1,3}(1-x) \right]. \tag{23}$$

The coefficient $\phi_2(x)$ was first calculated in another way in reference [2]. In order to estimate the accuracy of the fifth order solution I have calculated numerically the next coefficient of the Jadach–Ward series. The ratio of the fifth to the sixth order perturbative solutions of the GL equation using the Jadach–Ward *ad hoc* exponentiation prescription is shown in Fig.1. One can see that the accuracy of the fifth order solution is better than $1.6 \cdot 10^{-7}$ in the hard limit and better than $1 \cdot 10^{-8}$ in the soft one ($x \geq 0.8$).
In Fig.2 I compare existing third order solutions, exponentiated by Kuraev–Fadin (KF) [8] (for the explicit third order result see [2]) and Jadach–Ward (JW) prescriptions. Both solutions are normalized to the fifth order solution according to the JW exponentiation procedure. The comparison shows that the JW prescription is closer to the exact result than the KF one, especially in the soft limit.

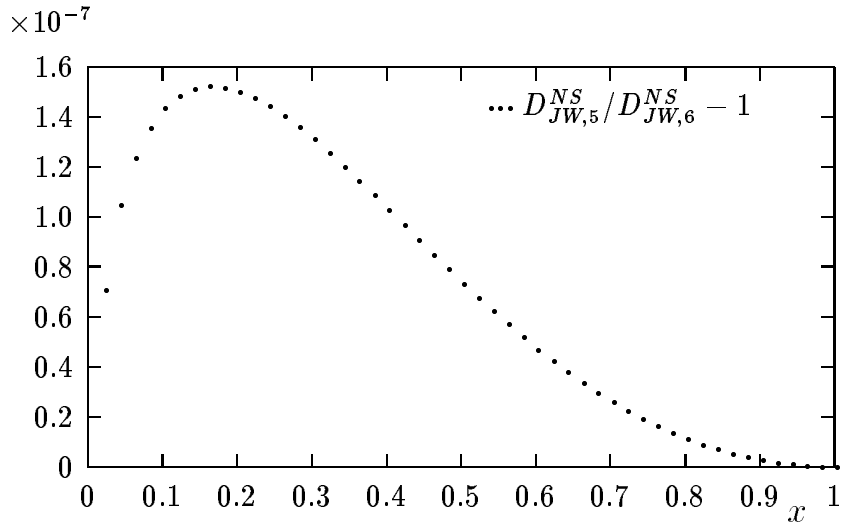

**Figure 1:** The ratio of the fifth ($D^{NS}_{JW,5}$) and sixth ($D^{NS}_{JW,6}$) order exponentiated solutions of the Jadach–Ward (JW) type for the LL non singlet electron structure function. The calculation was done for $\beta \simeq 0.11$ i.e. $\sqrt{s} = 92\text{GeV}$ for the LEP case.

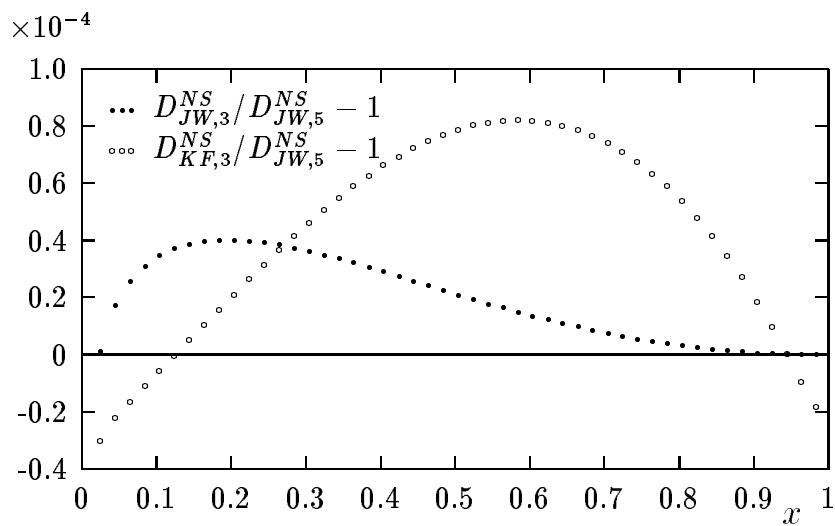

**Figure 2:** The ratio of the third order approximate solutions of the KF (open circles) and the JW (full circles) types to the fifth order solution of the JW type. The calculation was done for $\beta \simeq 0.11$ i.e. $\sqrt{s} = 92\text{GeV}$ for the LEP case.

In Fig.3 I present all the calculated coefficients of the Jadach–Ward series and the next three ($\phi_5$, $\phi_6$, $\phi_7$) which I have determined numerically. I have parametrized these last three coefficients using the following function:

$$\phi_i(x) = \left(-1 + (1-x)\frac{p_i(1) + p_i(2)x + p_i(3)x^2 + p_i(4)x^3 + p_i(5)x^4}{p_i(6) + p_i(7)x + p_i(8)x^2 + p_i(9)x^3 + p_i(10)x^4} + p_i(11)\ln x\right)\phi_{i-1}(x) \tag{24}$$

Parameters $p_i$ are collected in Table 1. The accuracy of the functions $\phi_i(x)$ for $i =$5, 6, 7, is better than $5 \cdot 10^{-6}$, in the range $0.01 \leq x \leq 1$.

**Table 1:** Numerical values of the parameters $p_i$ in equation (24).

| i | 5 | 6 | 7 |
|---|---|---|---|
| $p_i(1)$ | 0.641 664 028 | 0.003 843 748 | -0.000 103 279 |
| $p_i(2)$ | -0.619 988 501 | 0.006 263 836 | -0.002 208 024 |
| $p_i(3)$ | 0.224 513 784 | -0.033 976 652 | -0.061 516 549 |
| $p_i(4)$ | -1.589 189 529 | -0.210 926 712 | 0.130 086 362 |
| $p_i(5)$ | -2.087 939 501 | 0.156 033 173 | 0.759 156 644 |
| $p_i(6)$ | 0.227 220 744 | 0.010 454 907 | 0.000 757 690 |
| $p_i(7)$ | 110.025 192 261 | -2.389 885 902 | -0.249 451 116 |
| $p_i(8)$ | 68.255 920 410 | -35.333 641 052 | -7.883 694 649 |
| $p_i(9)$ | 260.508 605 957 | -30.422 819 138 | -264.039 489 746 |
| $p_i(10)$ | 195.836 334 229 | -204.567 718 506 | -1080.647 705 078 |
| $p_i(11)$ | -0.005 347 746 | 0.000 187 368 | -0.000 658 672 |

The shape of the approximate function reflects the fact that the fraction of the subsequent coefficients is very close to $-1$ (the following coefficients are of the opposite sign) in a wide $x$-range and the singularity of the $i$-th function for $x \to 0$ is of one power of the logarithm stronger than the $(i-1)$-th one.

From Fig.3 we see that the coefficients of the JW series appear to converge to two different functions: one for the odd and another one for the even coefficients. These limits, however, have an interesting property: they are very symmetric with respect to the x axis ($\phi_i(x) \equiv 0$). The $x$-range ($x_0 < x < 1$) where the symmetry follows with good accuracy increases with increasing coefficient function index.

In order to test how close the subsequent coefficient functions of the JW series are we present in Fig.4 the modules of their fractions. One can see that the coefficients $\phi_5, \phi_6$ and $\phi_7$ are very close to each other in the soft $x$-range (for $x > 0.5$ the difference is of the order of 0.01%, and for $x > 0.01$ the difference is up to a few percent).

The next coefficient, $\phi_8$, is not drawn in Fig.3 and Fig.4 because the difference between them and $\phi_7$ is smaller than 0.1% in the hard range and smaller than 0.01% in the soft one and would be not seen in the figures.

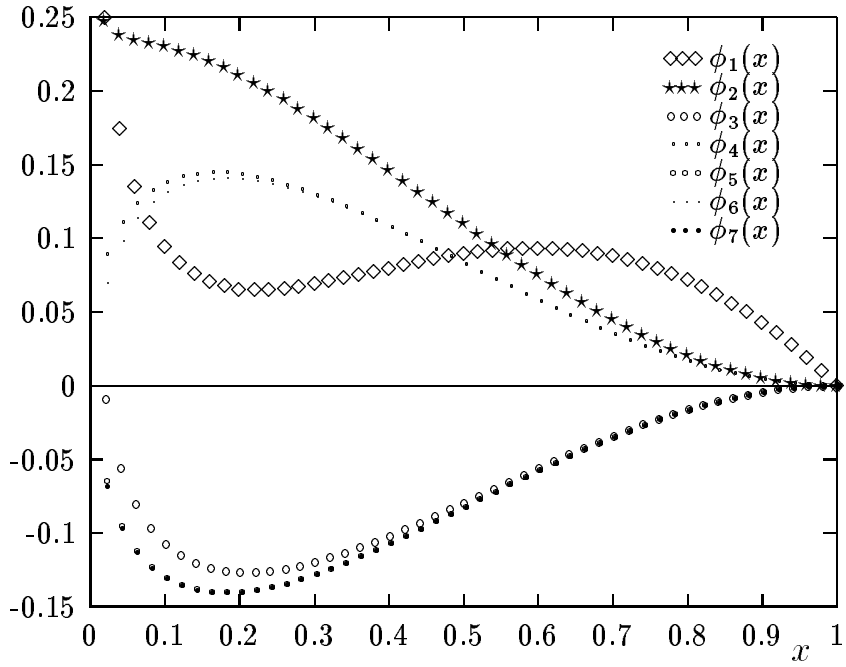

**Figure 3:** Comparison of all obtained (analytically and numerically) coefficients of the JW series.

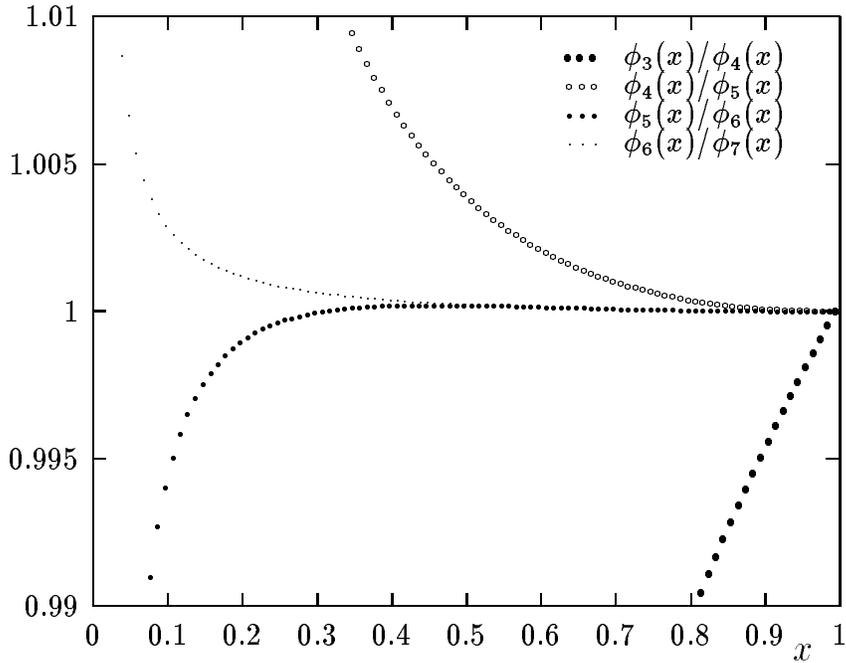

**Figure 4:** Comparison of the fractions of modules of the coefficients of the JW series.

# 3 Conclusions

Summarizing, we now have analytical, fifth order, $\mathcal{O}(\beta^5)$, corrections to the non–singlet electron LL structure function. The result was obtained according to the *ad hoc* exponentiation prescription of the Jadach–Ward type. The accuracy of the new solution is of the order of $10^{-7}$ in the hard limit and of $10^{-8}$ in the soft limit. The comparison with the existing third order solutions shows that the Jadach–Ward exponentiation provides an especially good aproximation. An approximate solution up to the eight order in $\beta$ (with numerical accuracy of the order of $10^{-6}$ in the range $0.01 < x < 1$) was also proposed. The higher order coefficient functions seem to be very close to $\phi_7$ in modulus especially in the soft limit. The subsequent coefficients are of the opposite sign in the wide x–range and seem to converge to two very symmetriclly placed limit functions: one for the odd and another one for the even coefficients.

# Acknowledgements

I would like to thank M.Jeżabek for useful discussions and critical reading of the manuscript.

# Appendix A

Generalized Nielsen's polilogarithms are defined as:

$$S_{n,m}(x) = \frac{(-1)^{n+m-1}}{(n-1)!\,m!} \int_0^1 \frac{dt}{t} \ln^{n-1} t \ln^m(1-xt) \tag{A.1}$$

$S_{n,m}(y)$, defined only for positive integers $n$ and $m$, is real for real $y \leq 1$. From the definition of $S_{n,m}(y)$ one can find its derivative and integral:

$$\frac{d}{dy} S_{n,m}(y) = \frac{1}{y} S_{n-1,m}(y) \tag{A.2}$$

$$\int_0^y \frac{dx}{x} S_{n,m}(x) = S_{n+1,m}(y) \tag{A.3}$$

In particular:

$$\frac{d}{dy} Li_2(y) = -\frac{\ln(1-y)}{y} \tag{A.4}$$

$$\frac{d}{dy} Li_3(y) = \frac{Li_2(y)}{y} \tag{A.5}$$

$$\frac{d}{dy} S_{1,2}(y) = \frac{\ln^2(1-y)}{2y} \tag{A.6}$$

where $Li_n(y) \equiv S_{n-1,1}(y)$. Some of relations between polylogarithms of different arguments:

$$Li_2(1-y) = -Li_2(y) - \ln(y)\ln(1-y) + \zeta(2) \tag{A.7}$$

$$Li_2(-\frac{y}{1-y}) = -Li_2(y) - \frac{1}{2}\ln^2(1-y) \tag{A.8}$$

$$Li_3(1-y) = -S_{1,2}(y) - \ln(1-y) Li_2(y)$$
$$\qquad - \frac{1}{2}\ln(y)\ln^2(1-y) + \zeta(2)\ln(1-y) + \zeta(3) \tag{A.9}$$

$$Li_3(-\frac{y}{1-y}) = S_{1,2}(y) - Li_3(y) + \ln(1-y)Li_2(y) + \frac{1}{6}\ln^3(1-y) \tag{A.10}$$

$$S_{1,2}(1-y) = -Li_3(y) + \ln(y)Li_2(y) + \frac{1}{2}\ln(1-y)\ln^2(y) + \zeta(3) \tag{A.11}$$

$$S_{1,2}(-\frac{y}{1-y}) = S_{1,2}(y) - \frac{1}{6}\ln^3(1-y) \tag{A.12}$$

where $\zeta(n) \equiv Li_n(1)$.

Some of definite integrals used in the calculations:

$$\int_0^1 \frac{\ln(c+ey)}{a+by} dy = \frac{1}{b}\left[\ln\left(\frac{bc-ae}{b}\right)\ln\left(\frac{a+b}{a}\right) - Li_2\left(e\frac{a+b}{ae-bc}\right) + Li_2\left(\frac{ae}{ae-bc}\right)\right] \tag{A.13}$$

$$\int_0^1 \frac{\ln(1-y)\ln(1-ay)}{y} dy = S_{1,2}(a) + Li_3(a) \tag{A.14}$$

$$\int_0^1 \frac{\ln(y)\ln(1-ay)}{1-y} dy = 2S_{1,2}(a) - Li_3(a) + \ln(1-a)[Li_2(a) - \zeta(2)] \tag{A.15}$$

$$\int_x^1 \frac{Li_2(1-y) - Li_2(1-x)}{y-x} dy = -S_{1,2}(1-x) - Li_3(1-x) \tag{A.16}$$

$$\int_x^1 \frac{Li_3(1-y) - Li_3(1-x)}{y-x} dy = -S_{2,2}(1-x) - Li_4(1-x) \tag{A.17}$$

$$\int_x^1 \frac{S_{1,2}(1-y) - S_{1,2}(1-x)}{y-x} dy = -2S_{1,3}(1-x) + S_{2,2}(1-x) - \frac{1}{2}Li_2^2(1-x) \tag{A.18}$$

More detailed information about polylogarithms and associated integrals can be found in [5],[6].